\documentclass[twocolumn,showpacs,preprintnumbers,amsmath,amssymb,prl,epsf]{revtex4}
\usepackage{graphicx}
\begin{document}
\title{Tailoring polarization entanglement in anisotropy-compensated SPDC}
\author{Giorgio Brida$^{1}$, Marco Genovese$^{1}$, Maria~V.~Chekhova$^{2}$, and Leonid~A.~Krivitsky$^{1,3}$}

\affiliation{$^{1}$Istituto Nazionale di Ricerca Metrologica, Strada delle
Cacce 91, 10135 Torino, Italy}
    \affiliation{$^{2}$Department of Physics, M.V.Lomonosov Moscow State
University
Leninskie Gory, 119992 Moscow, Russia}
\affiliation{$^{3}$ Institut fur Optik, Information und Photonik
Max-Planck Forschungsgruppe,
Universitaet Erlangen-Nuernberg\\
Guenther-Scharowsky-Str.1/Bau 24,
91058 Erlangen, Germany}

\begin{abstract}

\begin{center}\parbox{14.5cm}
{We show that the angular spectrum of type-II SPDC contains a
continuum of maximally entangled states. For the realization of a bright source of entangled two-photon states, a standard technique is to compensate for the transverse walk-off in the production crystal. We demonstrate that polarization entangled states can be effectively engineered by generalizing this approach. Our method allows to considerably enrich the set of generated entangled states with a controlled tailoring them.}
\end{center}
\end{abstract}
\pacs{ 03.67.Hk, 42.50.Dv, 42.62.Eh}
 \maketitle \narrowtext
\vspace{-10mm}

During the last decade, entanglement, initially a "paradoxical"
property of quantum systems being of purely theoretical interest,
became a fundamental resource promising interesting technological
applications, such as quantum communication or quantum
computation~\cite{NC,ol}, quantum metrology~\cite{las}, quantum
imaging~\cite{qi}, etc.

The most well-developed source of entangled states is Spontaneous
Parametric Down Conversion (SPDC), in which pairs of photons (biphotons) are
produced by the "decay" of pump photons inside a non-linear crystal.
There are several alternative schemes for generating polarization-entangled states ~\cite{Genovese}, based on type-II phase matching (producing orthogonally polarized photons)~\cite{au,Kwiat_rings,Branning}, and on type-I phase matching (producing photons of
identical polarization) ~\cite{Kwiat,Burlak,nos,mataloni}.

In the present paper we consider the most straightforward way to generate entangled states exploiting collinear type II SPDC. In this case the two directions where the emission cones of
different polarization (H,V) are superimposed for degenerate
photons coincide: since the two amplitudes add coherently one would generate
a $|\Psi^+\rangle={|HV\rangle+|VH\rangle}$ state. Nevertheless, a careful analysis shows that the birefringence of the non-linear crystal is responsible for temporal (\emph{longitudinal walk-off}) and spatial (\emph{transverse walk-off}) distinguishability of orthogonally polarized photons that lead to a reduced visibility in various interference experiments ~\cite{Kwiat_rings,bennink}. Thus, in order to improve the performance of SPDC based sources, an elegant technique was suggested for the compensation of the walk-off effect ~\cite{Kwiat_rings, kurt, Altepeter}. The most common solution here is to use a half-wave plate and a properly orientated additional birefringent crystal placed after the production one. Another possible way is to perform strong spectral and spatial selection by a narrow-band interference filters and restricted pinholes, for the price of considerable reduction of the count rate.

Despite the fact that the compensation technique mentioned above became a common tool in most of experiments exploiting type II SPDC, recently we have shown that the actual presence of the walk-off gives rise to an interesting effect ~\cite{bigPRAarxiv}. Namely, it has been shown that a continuous set of different maximally
entangled states of the form $| \Psi\rangle\sim{|HV \rangle +e^{i\phi} |VH \rangle}$ is generated within the line-shape with continuously varying value of the phase $\phi$, which can be chosen by means of frequency or angular selection~\cite{Brida}. In this paper we present a general approach to characterization of spatial-polarization entangled states produced via type II SPDC. We demonstrate that polarization entanglement can be effectively manipulated within the line-shape of SPDC by various birefringent materials.

%In our opinion this is a very important result since it should be always carefully considered whenever using SPDC entanglement for various applications.

Let us consider a biphoton produced by pumping a type-II
crystal with a cw-laser in the regime when signal and idler
photons with wavevectors $\bold k_s$ and $\bold{k_i}$, respectively, are close to degeneracy in frequencies and propagate nearly collinear to the pump wavevector $\bold{k_p}$ (see, e.g.,~\cite{BigDNK}). Further we assume that diameter of the pump beam is much larger than the transverse walk-off of down-converted photons given by $L\tan\theta$, where $\theta$ is the typical angle of scattering and $L$, the length of the crystal. In this case exact momentum conservation can be assumed in the transverse plane (orthogonal to $\bold{k_p}$). At the same time due to the  finite length of SPDC crystal the phase-matching condition is weakened in the longitudinal direction (along $\bold{k_p}$) and SPDC occurs also for non-perfectly phase-matched modes. Expanding the longitudinal mismatch $\bold{\Delta}(\bold k_s, \bold{k_i})$ around the collinear frequency-degenerate point, up to linear terms in frequency ($\Omega$) and angular ($\theta$) offsets, we obtain

\begin{equation}
\bold{\Delta}(\bold k_s, \bold{k_i})\equiv\bold k_p-\bold k_s-\bold k_i\simeq D\Omega+B\theta, \nonumber
\end{equation}
where $D\equiv\frac{dk_{e}}{d\Omega}-\frac{dk_o}{d\Omega},
B\equiv\frac{dk_{e}}{d\theta}$ and $k_e$, $k_o$ stand for the wavevectors of extraordinary (V-polarized) and ordinary (H-polarized) waves, respectively. Therefore in case of exact frequency degeneracy ($\Omega$=0) the state vector takes the form ~\cite{bigPRAarxiv}:
\begin{eqnarray}
|\Psi\rangle=\int_0^{\infty}{d\theta}\hbox{sinc}
\left(\frac{BL\theta}{2} \right)[|H_{-\theta}
V_{\theta}\rangle
%\nonumber\\
+e^{i\phi}|H_{\theta}V_{-\theta}\rangle
]
\label{IIang}
\end{eqnarray}
where the $\hbox{sinc}$ function is defined as $\hbox{sinc}(x)\equiv\frac{\sin
x}{x}$ and the relative phase $\phi=BL\theta$ is acquired for V-polarized photons propagating at angles $\theta$ and $-\theta$ due to the inequality of their refractive indexes $n_{e}(\theta)\neq{n_{e}(-\theta)}$. From (\ref{IIang}) it follows that because of the dependence of the relative phase $\phi$ on the scattering angle $\theta$, different entangled states are coherently generated within the line-shape of SPDC. For instance, the state generated at exact collinear direction, i.e., at $\theta=0$ is a factorable one, but it can be turned into the Bell state $\left|\Psi^+\right>$ by means of a 50/50 beamsplitter. At the same time, at angular offset $\theta=\pm\pi/BL$, the state already becomes another Bell one, namely the singlet state $\left|\Psi^-\right>$. At intermediate mismatch values, various maximally entangled states $|H_{-\theta}V_{\theta}\rangle+e^{i\phi}|H_{\theta}V_{-\theta}\rangle$ are generated with phases $\phi$ ranging from $0$ to $\pi$.

Hereafter we demonstrate that the "fine" polarization structure can be effectively
manipulated by reducing (compensating) and increasing (anti-compensating) the anisotropy of the production crystal. Let us consider that after elimination of the pump a piece of birefringent crystal is placed into the biphoton beam. In this case the relative phase acquired by the components of the state (\ref{IIang}) is given by the effective value $\phi+\phi'$, where $\phi'$ is a contribution acquired due to the presence of the additional crystal ~\cite{Altepeter}. This opens a possibility to engineer the two-photon state (\ref{IIang}) by introducing an additional phase $\phi'$. As we will show, in a certain case the generated state may have a uniform phase distribution within the spectrum, in contrast to the cases when the variation of the state becomes very rapid.

To demonstrate this, let us first consider the common method of transverse walk-off compensation, that is achieved by placing, after eliminating the pump, a second identical crystal but of a half length with an optical axis rotated at $180^{\circ}$. It can be easily shown that an additional phase $\phi'$ acquired by a pair passing through the compensator is exactly opposite to the phase $\phi$ which comes from the production crystal. Therefore $\phi'=-\phi$ and as a result the state does not depend on the scattering angle and becomes the Bell $\left|\Psi^+\right>$ state uniform all over the spectral line-shape:
\begin{eqnarray}
|\Psi\rangle\sim|H_{-\theta}
V_{\theta}\rangle
%\nonumber\\
+|H_{\theta}V_{-\theta}
\rangle.
\label{psicomp}
\end{eqnarray}
Thus the natural "fine" structure of the state (\ref{IIang}) can be washed out by placing a single crystal.

It is also interesting to consider the case where the effective phase changes more rapidly with an angle $\theta$ and consequently the state becomes much more inhomogeneous within the line-shape. To demonstrate this, it is sufficient to modify the "compensation" configuration described above by aligning an optical axis of the additional crystal in the same direction as the axis of the production one ~\cite{longitudinal}. In this case the effective phase is $\phi+\phi'=2\phi$ and the state changes twice more frequently with respect to the configuration without any additional crystal:
\begin{eqnarray}
|\Psi\rangle\sim|H_{-\theta}
V_{\theta}\rangle
%\nonumber\\
+e^{2iBL\theta}|H_{\theta}V_{-\theta}\rangle
\label{anticomp}.
\end{eqnarray}
It is obvious that the "anti-compensation" configuration is not interesting from the view point of efficient generation, since the line-shape consists of too many different polarization states. However this scheme has an important advantage if one is interested in generating many different entangled states from a single crystal. For instance, selecting angles at which $e^{2iBL\theta}=-1$, such as $\theta_1=\pm\pi/2BL$ and $\theta_2=\pm3\pi/2BL$ two singlet Bell states $|\Psi^{-}_{1}\rangle=|H_{-\theta_1}V_{\theta_1}\rangle-|H_{\theta_1}V_{-\theta_1}\rangle
$ and $|\Psi^{-}_{2}\rangle=|H_{-\theta_2}V_{\theta_2}\rangle-|H_{\theta_2}V_{-\theta_2}\rangle$ will be simultaneously emitted from a single crystal with the amplitudes proportional to $\hbox{sinc}\left(\frac{\pi}{4} \right)\approx0.9$ and $\hbox{sinc}\left(\frac{3\pi}{4} \right)\approx0.3$, respectively. It is also evident that, the structure of generated states can be easily enriched using a longer additional crystal, however this would require a very precise selection of angular modes since the phase $\phi$ will change very rapidly with the variation of $\theta$.

The  experimental setup used for this work is shown in Fig.1. A cw-Ar laser at $351$ nm pumped a $1$ mm type-II BBO crystal cut for collinear operation at degenerate wavelength $702$ nm. After eliminating the pump beam by a highly reflective UV-mirror, compensation and anti-compensation of transverse walk-off was studied at different orientations of $0.5$ mm BBO crystal, cut in the same way as the production one. After that SPDC radiation was split using a $50/50$ non-polarizing beamsplitter and a Glan prism was put into each
output port. Angular selection was performed in one output port of
the beamsplitter by using a lens with the focal length $500$ mm,
and an avalanche photodiode detector (Perkin-Elmer SPCM), whose sensitive element has
 size of $200\mu\times 200\mu$, placed in its focal plane (preceded by
an interference filter with FWHM=$3$ nm, centered at the degenerate
wavelength $702$ nm). The other arm of the setup (arm 2) was designed  to collect all
angular width of SPDC spectrum corresponding to the degenerate
frequency. For this purpose, we imaged the pump beam waist in the crystal ($1.5$
mm) onto the sensitive area of the avalanche photodiode  with a
demagnification of $1:6$. This was obtained by using an objective
lens with a large numerical aperture and with the focal length
$F=100$ mm. The angular width of SPDC after the lens, although
increased $6$ times, was still within the acceptance angle of the
photodiode. In order to eliminate the effect of the longitudinal walk-off and to prevent a significant contribution of accidental coincidences, an interference filter with FWHM=$1$ nm and a vertical slit of $2$ mm width were placed before the detector. Finally, coincidences of the two detectors were measured by addressing the output of one detector as start and the other as stop of a Time to Amplitude Converter. The scanning was performed in the plane of the optic axis. This way we studied the corresponding dependence of the coincidence
counting rate on the angle selected by the pinhole in arm 1 of the
setup. In order to characterize polarization entanglement we studied the angular dependence of coincidences for the $(45^{\circ},45^{\circ})$ and $(45^{\circ},-45^{\circ})$ settings of the Glan prisms.

\begin{figure}
\includegraphics[width=0.5\textwidth]{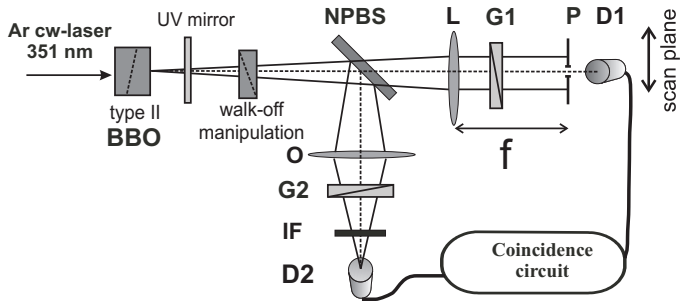}
\caption{Scheme of the experimental setup. SPDC is produced in 1 mm type II BBO crystal; walk-off manipulation performed with 0.5 mm BBO at different orientations; the pump is eliminated by a highly-reflective UV mirror;
\textbf{NPBS}, a 50/50 non-polarizing BS; \textbf{G1} and \textbf{G2},
Glan prisms; \textbf{D1}, \textbf{D2}, single-photon counting modules. Angular
selection is performed in arm 1 by a pinhole \textbf{P} translated in the plane of the optic axis and preceded by a lens \textbf{L} with focal length
$500$ mm. In arm 2, all angles are collected by means of an
objective-lens \textbf{O} with focal length $100$ mm. Narrow-band interference
filter \textbf{IF} is used to avoid the contribution of accidental coincidences and the effect of longitudinal walk-off.}
\end{figure}

For the state (\ref{IIang}) generated in our set-up the dependence of the coincidences counting rate on the orientations of Glan prisms $\Theta_1$ and $\Theta_2$ has the form
\begin{eqnarray}
R_c(\theta)=\hbox{sinc}^2 \left(\frac{B\theta L}{2} \right)[\sin^2
\left(\Theta_1+\Theta_2 \right)\cos^2\left(\frac{\phi}{2}
\right) \nonumber
\\
+\sin^2(\Theta_1-\Theta_2)\sin^2 \left(\frac{\phi}{2}\right)].
\label{coincidences}
\end{eqnarray}
First, we studied the angular coincidences distributions in the absence of the additional BBO crystal, when the relative phase in (\ref{IIang}) is $\phi=B{\theta}L$. The corresponding coincidences distributions versus scanned angle $\theta$ at $\Theta_1=45^{\circ},\Theta_2=\pm45^{\circ}$ presented in Fig.2a. Observation of an oscillating behavior of coincidences at given orientations of the polarizers shows that different entangled states are generated within the line-shape. Namely, the state generated at the center of the angular line-shape is $\left|\Psi^+\right>$ (marked with vertical solid bar), whilst at the angles $\pm0.0055$ rad (marked with two vertical dotted bars) the state $\left|\Psi^-\right>$ is produced.

\begin{figure}
\includegraphics[height=4.5cm]{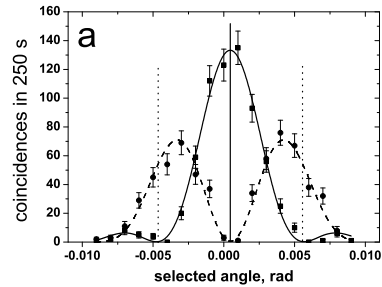}
\includegraphics[height=4.5cm]{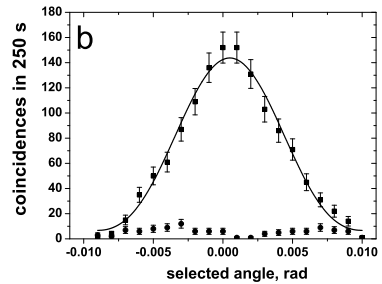}
\includegraphics[height=4.5cm]{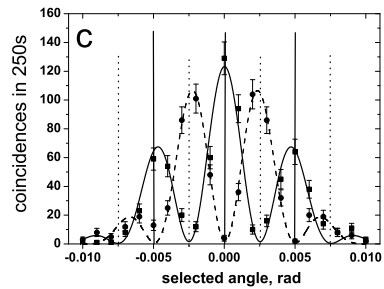}
\caption{Experimental dependence of the coincidence counting rate
on the angle selected by the pinhole in arm 1 of the setup for two cases: $\Theta_1=\Theta_2=45^\circ$
(squares, solid line) and $\Theta_1=45^\circ, \Theta_2=-45^\circ$ (circles, dashed line). (a)- without an additional crystal; (b)- with an additional crystal in "compensation" and (c) in "anti-compensation" configurations. Vertical bars indicate the angles at which $\left|\Psi^+\right>$ (solid) and $\left|\Psi^-\right>$ (dotted) states are generated. Lines represent the fit of experimental data using (\ref{coincidences}).}
\end{figure}

After that we studied the effect of spatial compensation, placing an additional BBO crystal of $0.5$ mm with the opposite optic axis orientation. In this case the effective phase in (\ref{IIang}) is equal to zero and $\left|\Psi^+\right>$ state is generated all over the line-shape. This is indeed confirmed by the observed coincidences distribution in Fig.2b. Namely, the coincidence counting rate at ($\Theta_1=45^{\circ},\Theta_2=-45^{\circ}$) is strongly suppressed for all angles, whilst the one at ($\Theta_1=45^{\circ},\Theta_2=45^{\circ}$) takes the shape of the spectrum envelope.

We also studied the visibility of polarization interference as a function of the acceptance angle defined by the opening diameter of the pinhole in arm 1. Once the pinhole was centered at $\theta=0$, for every value of pinhole diameter we measured the number of coincidences $C(\Theta1;\Theta2)$ at certain orientations of the polarizers $\Theta1$ and $\Theta2$ and calculated the visibility as $V=|\frac{C(45^\circ;45^\circ)-C(45^\circ;-45^\circ)}{C(45^\circ;45^\circ)+C(45^\circ;-45^\circ)}|$. The experimental results without and with compensation are shown in Fig.3. The fact that in the absence of the compensation crystal the visibility decreases with an increase in the acceptance angle can be intuitively understood as follows: since a continuous set of entangled states is generated in this case, a wide open pinhole transmits a lot of spatial modes every one of which carries a different polarization state. Therefore, at a wide open pinhole one observes a completely mixed state and consequently its visibility is close to zero. In contrast, for the compensated configuration the state is uniform all over the line-shape and opening the pinhole does not spoil the interference keeping visibility at initially high level, despite the increase in the number of coincidences.

\begin{figure}
\includegraphics[width=0.5\textwidth]{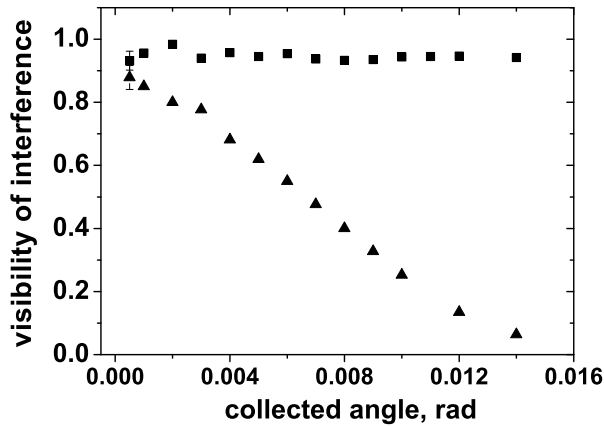}
\caption{Experimental dependence of the visibility of polarization interference on the scattering angle collected in experiment for two cases: without an additional crystal (triangles) and with additional crystal in "compensating" configuration (squares).}
\end{figure}

The final set of experiments was carried out when we turned the additional BBO crystal in "anti-compensation" configuration. In this case the effective phase in (\ref{IIang}) is given by $\phi+\phi'=2DL\theta$ and according to the discussion above one should observe modulation twice as frequent as the one obtained without an additional crystal. This is indeed confirmed from our experimental results presented in Fig.2c. As it was mentioned above the  "anti-compensation" configuration can be quite useful when one requires multiple entangled states generated from a single crystal. As it follows from (\ref{anticomp}) and our experimental results (see Fig.2c), the $\left|\Psi^+\right>$ state is generated at the angles where $2DL\theta=2{\pi}N, N=0,1..$ and these points are marked with vertical solid bars. Furthermore, at angles where $2DL\theta={\pi}(2N+1)$ the $\left|\Psi^-\right>$ state is generated and these values are marked with vertical dotted bars. It is worth mentioning that of course anti-compensation configuration introduces further possibilities to manipulate with the line-shape of SPDC by varying the parameters of the additional crystal such as its orientation and length.

In conclusion, we have reported a phenomenon that, to the best of our knowledge, was not considered before, namely
the presence inside the SPDC angular line-shape of a continuum of maximally
entangled states with varying phase between its components. We show that this
"fine"structure of SPDC natural line-shape is defined  by the transverse walk-off,
usually compensated without proper analysis.
We have shown that the spatial structure of the two-photon state can be effectively
manipulated by inserting additional birefringent materials depending on the
desired applications. This result, since it allows a tailoring of the produced state, is, in our opinion, of utmost importance for the
search of high-efficiency sources of entangled states and thus for
quantum information and related fields (quantum metrology, imaging,
etc.).

This work was supported in part by the
joint grant RFBR-Piedmont 07-02-91581-ASP, by italian minister of
research (PRIN 2005023443-002) and by Regione Piemonte (E14). One of us (L.A.K.)
acknowledges the support of the Alexander Von Humboldt foundation and another one (M.V.C.) the support of DFG Mercator fellowship.


\begin{references}
\bibitem{longitudinal} In this case additional longitudinal walk-off should be compensated, either by an additional
compensation element or by a narrow frequency filter.
\end{references}
\end{document}